\documentclass{elsart}
\usepackage{epsfig}
\usepackage{amsfonts}
\begin{document}
\begin{frontmatter}

\title{Nonextensive scalar field theories and dark energy
models}

\author{Christian Beck}

\address{School of Mathematical Sciences, Queen Mary, University of
London, Mile End Road, London E1 4NS, UK}

\begin{abstract}
Current astronomical measurements indicate that approximately
73$\%$ of the universe is made up of dark energy. Stochastically
quantized self-interacting scalar fields can serve as suitable
models to generate dark energy. We study a particular model where
the scalar field theory underlying dark energy exhibits strongest
possible chaotic behaviour.
It is shown that the fluctuating chaotic
field momenta obey a generalized (nonextensive) statistical
mechanics, with an entropic index $q$ given by $q=3$, respectively
$q=-1$ if the escort formalism is used.
\end{abstract}

\end{frontmatter}

\section{Introduction}
The formalism of nonextensive statistical mechanics has been
developed as a useful tool to statistically describe complex
systems \cite{nonex,mendes,abe}. The concept is based on
generalizing the Shannon entropy $S_1$ to more general information
measures, such as the Tsallis entropies $S_q$\cite{nonex}, or other
generalized information measures as well\cite{cos2,naudts,souza}, and
doing statistical mechanics based on these more general
information measures. Often, the formalism is relevant to
nonequilibrium systems with spatio-temporal parameter
fluctuations, which exhibit a superposition of different
statistics on different time scales (in short, a
'superstatistics') \cite{souza,super,euro,reynolds}. In recent
years, several physical applications have been pointed out for the
formalism with $q\not= 1$, among them being turbulent
flows\cite{euro,reynolds}, pattern forming
systems\cite{boden}, heavy ion collisions\cite{ion}, cosmic
rays\cite{cos1,cos3,cos2}, cosmology\cite{pennini},
econophysics\cite{eco}, and biological
systems \cite{bio}.

In this paper we want to point out yet another potential
application of the nonextensive formalism, namely to
stochastically quantized dark energy models. There is growing
observational evidence that the universe is presently dominated by
some unknown form of vacuum energy with equation of state close to
$w=-1$, so-called dark energy.
Current astronomical data indicate that there is approximately 73
\% dark energy, 23 \% dark matter, and 4\% ordinary matter in our
universe\cite{dark1,dark2}. The nature and origin of the
dominating dark energy component is not understood, and many
different models are currently being discussed
\cite{cosco1,cosco2,astro}.

The simplest way to generate dark energy is via a self-interacting
scalar field. The dark energy density is then essentially given by
the potential of the field, in either a static way as for a true
cosmological constant or in a dynamical way as
in quintessence models\cite{cosco1,cosco2}. In this paper we
consider a model for dark energy which associates dark energy with
self-interacting scalar fields corresponding to a
$\varphi^4$-theory, which is second quantized \cite{astro,book,physicad}. The
difference to other scalar field approaches is that our fields are
very strongly (rather than weakly) self-interacting, and that they
are second-quantized. We use as the relevant method to quantize
the scalar fields the stochastic quantization method introduced by
Parisi and Wu \cite{stoch1,stoch2}. In the fictitious time
variable of this approach, the fields behave in a deterministic
chaotic way. Our physical interpretation is to associate the
chaotic behaviour of the scalar fields with vacuum fluctuations
\cite{book,physicad}. When quantum mechanical averages are formed,
the fields generate quite smoothly distributed dark energy
\cite{astro}.



We will show that the probability density of the fluctuating
chaotic momenta in our model is given by the generalized canonical
distributions in the formalism of nonextensive statistical
mechanics. The relevant entropic index is $q=3$
(see also \cite{pennini}), or $q=-1$ if the
escort formalism is used. This means the dark energy component
behaves quite different from a system of particles described by
ordinary statistical mechanics.
The expectation of the scalar field potential plays the role of a
generalized thermodynamic potential in this setting. We will show
that a theory formulated in terms of escort distributions, i.e.\
$q=-1$, has many advantages as compared to $q=3$. We will
explicitly calculate quantities like the entropy $S_q$, the
generalized internal energy $U_q$ and the generalized free energy
$F_q$ for our model.


If our model is indeed the correct description of dark energy in
the universe, then this means that Boltzmann-Gibbs type of
statistical mechanics ($q=1$) is only relevant for a minority of
the contents of the universe (ordinary and dark matter, together
27 $\%$), whereas the dominating dark energy contents of the
universe (73 $\%$) is described by a different type of statistical
mechanics, with $q=-1$.

This paper is organized as follows. In section 2 we first recall
how to second-quantized a scalar field dynamics using stochastic
quantization. We then show that a chaotic dynamics can be obtained
if the potential is very strong. Our main example in section 3 is
a $\varphi^4$-theory leading to an effective dynamics given by
weakly coupled 3rd order Tchebyscheff maps. This dark energy model
exhibits strongest possible chaotic behavior. In section 4
we show how the
parameters have to be chosen in order to reproduce the currently
measured dark energy. Finally, in section 5 we embed the chaotic field
model into the formalism of nonextensive statistical mechanics
and calculate various thermodynamic quantities.


\section{Stochastic quantization of strongly self-inter\-acting scalar fields}

Let us consider a homogeneous self-interacting scalar field $\varphi$ in
Robertson-Walker metric. To second-quantize it, one can use the
Parisi-Wu approach of stochastic quantization
\cite{stoch1,stoch2}. This means one considers the following
stochastic differential equation:
\begin{equation} \frac{\partial}{\partial s}\varphi
=\ddot{\varphi} +3H\dot{\varphi} +V'(\varphi) +L(t,s) \label{sto}
\end{equation}
Here $H$ is the Hubble parameter, $V$ is the potential under
consideration, $t$ is physical time, and $s$ is an artificial
coordinate (called fictitious time) which is just introduced for
quantization purposes. $L(t,s)$ is Gaussian white noise,
$\delta$-correlated both in $t$ and $s$. The expectations with
respect to the stochastic process generated by this stochastic
differential equation for $s\to \infty$ are identical to the
quantum mechanical expectations of the field theory under
consideration.

We may discretize eq.~(\ref{sto}) using $ s = n\tau$ and $t = i
\delta $, where $n$ and $i$ are integers and $\tau$ is a
fictitious time lattice constant, $\delta$ is a physical time
lattice constant. The continuum limit requires $\tau \to 0$,
$\delta \to 0$, but we will later see that it makes
physical sense to keep small but finite lattice constants of the
order of the Planck length. We obtain
\begin{equation}
\frac{\varphi_{n+1}^i-\varphi_n^i}{\tau} = \frac{1}{\delta^2}
(\varphi_n^{i+1}-2\varphi_n^i+\varphi_n^{i-1}) +3\frac{H}{\delta}
(\varphi_n^i-\varphi_n^{i-1}) +V'(\varphi_n^i) + noise
\end{equation}
equivalent to
\begin{eqnarray}
\varphi_{n+1}^i&=& (1-\alpha)\left\{ \varphi_n^i
+\frac{\tau}{1-\alpha}
V'(\varphi_n^i)\right\}+3\frac{H\tau}{\delta} (\varphi_n^i-
\varphi_n^{i-1})+\frac{\alpha}{2}(\varphi_n^{i+1}+\varphi_n^{i-1})
\nonumber \\ \, &+& \tau\cdot noise,
\end{eqnarray}
where a dimensionless coupling constant $\alpha$ is introduced as
$\alpha:=\frac{2\tau}{\delta^2}$. We also introduce a
dimensionless field variable $\Phi_n^i$ by writing
$\varphi_n^i=\Phi_n^i p_{max}$, where $p_{max}$ is some (so far)
arbitrary energy scale. The discretized stochastically quantized
system is equivalent to a coupled map lattice
of the form
\begin{equation}
\Phi_{n+1}^i=(1-\alpha)T(\Phi_n^i)+\frac{3}{2}H\delta \alpha
(\Phi_n^i-\Phi_n^{i-1})+\frac{\alpha}{2}(\Phi_n^{i+1}+\Phi_n^{i-1})
+ \tau\cdot noise, \label{dyn}
\end{equation}
where the local map $T$ is given by
\begin{equation}
T(\Phi )=\Phi
+\frac{\tau}{p_{max}(1-\alpha)}V'(p_{max}\Phi).\label{map}
\end{equation}
Here the prime means $ '=\frac{\partial}{\partial
\varphi}=\frac{1}{p_{max}}\frac{\partial}{\partial \Phi}$. For
large $t$ (old universes) the term proportional to $H$ can be
neglected and one obtains a symmetric diffusively coupled map
lattice of the form
\begin{equation}
\Phi_{n+1}^i=(1-\alpha)T(\Phi_n^i)+\frac{\alpha}{2}(\Phi_n^{i+1}+\Phi_n^{i-1})
+\tau \; noise. \label{sym}
\end{equation}
Apparently the iteration of a coupled map lattice of the form
(\ref{sym}) with a given map $T$ has physical meaning: It means
that one is considering the second-quantized dynamics of a
self-interacting real scalar field $\varphi$ with a force $V'$
given by
\begin{equation}
V'(\varphi)=\frac{1-\alpha}{\tau} \left\{ -\varphi +p_{max}
T\left(\frac{\varphi}{p_{max}}\right) \right\},
\end{equation}
equaivalent to
\begin{equation}
V(\varphi) =\frac{1-\alpha}{\tau} \left\{ -\frac{1}{2} \varphi^2 +
p_{max}\int d\varphi \,T\left(\frac{\varphi}{p_{max}}\right)
\right\} + const .\label{pot}
\end{equation}

\section{Chaotic $\varphi^4$-theory}

It is interesting to see that our model can exhibit
chaotic behaviour for very strong forces 
$V'\sim \tau^{-1}$.
As a distinguished
example, 
let us consider the map
\begin{equation}
\Phi_{n+1}=T_{-3}(\Phi_n)=-4\Phi_n^3+3\Phi_n
\end{equation}
on the interval $\Phi\in [-1,1]$. $T_{-3}$ is the negative
third-order Tchebyscheff map, a strongly chaotic map
conjugated to a Bernoulli shift of 3 symbols. This map arises in the
scalar field dynamics (\ref{sym}) if
the scalar field potential is given by
\begin{equation}
V_{-3}(\varphi)=\frac{1-\alpha}{\tau}\left\{
\varphi^2-\frac{1}{p_{max}^2} \varphi^4\right\}+const,
\end{equation}
or, in terms of the dimensionless field $\Phi$,
\begin{equation}
V_{-3}(\varphi)=\frac{1-\alpha}{\tau} p_{max}^2 ( \Phi^2 -\Phi^4)
+ const.
\end{equation}
Apparently, this potential describes a field $\varphi$ that
rapidly fluctuates in fictitious time on some finite interval .
The small noise term in eq.~(\ref{sym}) can be neglected as
compared to the deterministic chaotic fluctuations of the field.

Of physical relevance are the expectations of suitable observables
with respect to the ergodic chaotic dynamics. For example, one can
calculate the expectation of the potential $V_{-3}$, which describes
the vacuum energy generated by this chaotic field
theory:
\begin{equation} \langle
V_{-3}(\varphi)\rangle =\frac{1-\alpha}{\tau} p_{max}^2 ( \langle
\Phi^2\rangle  -\langle \Phi^4\rangle ) + const
\end{equation}
For uncoupled Tchebyscheff maps ($\alpha=0$), expectations of any
observable $A$ can be evaluated as the ergodic average
\begin{equation}
\langle A \rangle = \int_{-1}^{+1} A(\Phi) p (\Phi) d\Phi,
\label{obs}
\end{equation}
with the natural invariant density $p(\Phi)$ being given by
\begin{equation}
p (\Phi) =\frac{1}{\pi\sqrt{1-\Phi^2}}. \label{mu}
\end{equation}
From eq.~(\ref{mu}) one obtains $\langle \Phi^2 \rangle
=\frac{1}{2}$ and $\langle \Phi^4 \rangle =\frac{3}{8}$, thus
\begin{equation} \langle V_{-3} (\varphi)
\rangle=\frac{1}{8}\frac{p_{max}^2}{\tau} +const.
\end{equation}

Alternatively, we may consider the positive Tchebyscheff map $T_3
(\Phi)=4\Phi^3-3\Phi$. This basically exhibits the same dynamics
as $T_{-3}$, up to a sign. Repeating the same calculation we
obtain
\begin{equation}
V_{3}(\varphi)=\frac{1-\alpha}{\tau} p_{max}^2 ( -2\Phi^2
+\Phi^4).
\end{equation}
For the expectation of the vacuum energy one gets
\begin{equation}
\langle V_{3}(\varphi)\rangle =\frac{1-\alpha}{\tau} p_{max}^2 (
-2\langle \Phi^2\rangle  +\langle \Phi^4\rangle ) + const,
\end{equation}
which for $\alpha=0$ reduces to
\begin{equation}
\langle V_{3} (\varphi) \rangle=-\frac{5}{8}\frac{p_{max}^2}{\tau}
+const.
\end{equation}
Symmetry considerations between $T_{-3}$ and $T_3$ suggest to take
the additive constant $const$ as $const=+\frac{1-\alpha}{\tau}
p_{max}^2 \frac{1}{2} \langle \Phi^2 \rangle$. In this case one
obtains the fully symmetric equation
\begin{equation}
\langle V_{\pm 3} (\varphi)\rangle =\pm
\frac{1-\alpha}{\tau}p_{max}^2 \left\{ -\frac{3}{2} \langle \Phi^2
\rangle +\langle \Phi^4 \rangle \right\},
\end{equation}
which for $\alpha\to 0$ reduces to
\begin{equation}
\langle V_{\pm 3} (\varphi)\rangle =\pm \frac{p_{max}^2}{\tau}
\left( -\frac{3}{8} \right). \label{xxx}
\end{equation}

The above vacuum energy
changes in a nontrivial way with the
coupling $\alpha$. Extensive numerical simulations were performed
in \cite{book,physicad}, and it was found that as a function of
$\alpha$ it has
several local minima that numerically coincide with running
standard model coupling constants, evaluated at the mass scales of
known fermions and bosons. This emphasizes the physical relevance
of the model. The role of chaotic fields in the universe could be
to fix and stabilize fundamental constants as local minima in the
dark energy landscape.

\section{Constructing a dark energy model}

Let us now fix the free parameters in our chaotic field model in
such a way that the currently experimentally measured properties
of dark energy come out correctly.

Firstly, the measured dark energy density is positive, not
negative. This means we have to choose the negative Tchebyscheff
map $T_{-3}$.

Next, consider the parameter $\tau$. It is the lattice constant of
fictitious time $s$ and has dimension $GeV^{-2}$. Ordinary
stochastic quantization based on Gaussian white noise requires the
continuum limit $\tau \to 0$. But since quantum field theory runs
into difficulties at the Planck scale $m_{Pl}$ and is expected to
be replaced by a more advanced theory at this scale, it is most
reasonable to take the small but finite value
\begin{equation}
\tau \sim m_{Pl}^{-2}\sim 10^{-38} GeV^{-2}. \label{tau}
\end{equation}

Next, the classical equation of state of dark
energy should be close to $w=-1$, as experimentally measured 
\cite{dark1,dark2}.
The classical equation of state is
defined by $w=\langle p \rangle/\langle \rho \rangle$, where $p$
is the pressure and $\rho$ the energy density. As shown in
\cite{astro}, this condition requires that the Tchebyscheff maps
are weakly coupled. $\alpha =0$ generates an equation of state
exactly given by $w=-1$, and $\alpha << 1$ generates an equation
of state $w\approx -1$.

Finally, the parameter $p_{max}$ can be determined from the
currently measured dark energy density \cite{dark1,dark2}
\begin{equation}
\rho_\varphi^{Obs} =(2.9\pm 0.3)\cdot 10^{-47} GeV^4. \label{oops}
\end{equation}
For $\alpha \approx 0$, eq.~(\ref{xxx}), (\ref{tau})
and (\ref{oops}) fix the
parameter $p_{max}$ as
\begin{equation}
p_{max}\sim 10^{-42} GeV.
\end{equation}
In static models, where the vacuum energy density does not change
with the expansion of the universe, the smallness of this
parameter is the famous cosmological constant problem
\cite{cosco1,cosco2}. In models with dynamically evolving dark
energy, the (current) smallness of this
parameter is related to the fact that the universe is old\cite{astro}.

\section{Generalized statistical mechanics}

The type of chaotic vacuum fluctuations considered here have
rather interesting properties, which can be described in the
language of statistical mechanics. For $\alpha = 0$ the natural
invariant measure describing the probability distribution of the
fluctuating variables $\Phi_n^i$ is given by eq.~(\ref{mu}),
and the measure factorizes at all lattice sites. The relevant
probability density can be regarded as a generalized canonical
distribution in non-extensive statistical mechanics \cite{nonex}.
As it is well known, one defines for a
dimensionless continuous random variable $X$ with probability density
$p(x)$ the
Tsallis entropies as
\begin{equation}
S_q=\frac{1}{q-1}(1-\int p(x)^qdx).\label{sq}
\end{equation}
Here $q$ is the entropic index. The Tsallis entropies contain the
Shannon entropy $S_1$ as a special case for $q\to 1$. Extremizing
$S_q$ subject to the constraint
\begin{equation}
\int p(x)E(x)dx =U \label{constraint}
\end{equation}
one ends up with $q$-generalized canonical distributions. These
are given by
\begin{equation}
p(x)\sim (1+(q-1)\beta E(x))^{-\frac{1}{q-1}}, \label{can}
\end{equation}
where $E$ is the energy associated with microstate $x$,
and $\beta$ is the inverse temperature. Of course,
for $q\to 1$ one obtains the usual Boltzmann factor $e^{-\beta
E}$.

Alternatively, one can work with the escort distributions, defined
by \cite{escort}
\begin{equation}
P(x)=\frac{p(x)^q}{\int p(x)^q dx}. \label{escort}
\end{equation}
If the energy constraint (\ref{constraint}) is implemented using the escort
distributions $P(x)$ rather than the original distribution $p(x)$,
one obtains generalized canonical distributions of the form
\begin{equation}
P(x)\sim (1+(q-1)\beta E(x))^{-\frac{q}{q-1}}. \label{Can}
\end{equation}
Again the limit $q\to 1$ yields ordinary Boltzmann factors
$e^{-\beta E}$.

Let us now apply this formalism to the chaotic fields $X=\Phi_n^i$. We
might identify $E=\frac{1}{2}m\Phi^2$ as a formal kinetic energy
associated with the chaotic fields. We then get by comparing
eq.~(\ref{mu}) and (\ref{can})
\begin{eqnarray}
q&=&3\\ \beta^{-1}&=&-m
\end{eqnarray}
Two problems arise: Firstly, the temperature that formally comes
out of this approach is negative. Secondly, and more seriously,
the Tsallis entropy of the distribution (\ref{mu})
as defined by the integral eq.~(\ref{sq}) does
not exist, since the integral $\int_{-1}^1 (1-x^2)^{-3/2}dx$
diverges.

However, it is remarkable that these problems do not occur if the
escort formalism \cite{mendes} is used. By comparing eq.~(\ref{mu}) and
(\ref{Can}) we obtain
\begin{eqnarray}
q&=& -1 \\ \beta^{-1} &=&m.
\end{eqnarray}
We obtain the result that our dark energy model behaves similar to
an ideal gas in the nonextensive formalism but with entropic index
$q=-1$, rather than $q=1$ as in ordinary
statistical mechanics. The 'velocity' $v$ is given by the
dimensionless variable $v=\Phi_n^i$, the 'energy' by the
non-relativistic formula $E=\frac{1}{2}mv^2$, and the inverse
temperature is $\beta =m^{-1}$. This nonextensive gas has the
special property that the temperature coincides with the mass of
the 'particles' considered. For nonzero $\alpha$ a similar
formalism remains valid, just that the effective energies $E$
become more complicated \cite{book}.

We can now evaluate all interesting thermodynamic properties of
the system using the escort formalism. Regarding the invariant
density (\ref{mu}) as an escort distribution $P(\Phi)$, the
original distribution is given by
\begin{equation}
p(\Phi)=\frac{2}{\pi}\sqrt{1-\Phi^2}\sim P(\Phi)^{1/q}. \label{pphi}
\end{equation}
For the Tsallis entropy of the chaotic fields we obtain from
eq.~(\ref{sq}) and (\ref{pphi})
\begin{equation}
S_{q}[p]=\frac{1}{4}(\pi^2 -2)=S_{q}[P] \;\;\;(q=-1)
\end{equation}
It is invariant under the transformation $p\to P$ (a distinguished property
of the entropic index $q=\pm 1$).

For the generalized internal energy we obtain
\begin{equation}
U_{q}[P]=\int_{-1}^1P(\phi)\frac{1}{2}m\Phi^2 =\frac{1}{4} m
\;\;\;(q=-1)
\end{equation}
and for the generalized free energy
\begin{equation}
F_q=U_q-TS_q=\frac{m}{4}(3-\pi^2) \;\;\;(q=-1).
\end{equation}


All expectations formed with the invariant measure can be regarded
as corresponding to escort expectations within the more general
type of thermodynamics that is relevant for our chaotically
evolving scalar fields. Whereas ordinary matter is described by a
statistical mechanics with $q=1$, the chaotic fields generating
dark energy are described by $q=-1$. The role of dark energy
is similar to that of a suitable thermodynamic potential 
in this
more general type of statistical mechanics. In general, one can
easily verify that the entropic indices $q=+ 1$ and $q=-1$ are
very distingushed cases: Only for these two cases the Tsallis entropy
of the escort distribution is equal to the Tsallis entropy of the
original distribution, for arbitrary distributions $p(x)$. 
If our model of dark energy is correct, then both types of statistical
mechanics
are realized in the universe.

\end{document}